\documentclass{ws-mpla}

\begin{document}

\markboth{R. Nelson and T. Mart}
{Kaon photoproduction on the nucleon with constrained parameters}

\catchline{}{}{}{}{}

\title{KAON PHOTOPRODUCTION ON THE NUCLEON WITH CONSTRAINED PARAMETERS}

\author{\footnotesize R. NELSON and T. MART}

\address{Departemen Fisika, FMIPA, Universitas Indonesia, 
Depok, 16424, Indonesia\\
tmart@fisika.ui.ac.id}

\maketitle


\begin{abstract}
The new experimental data of kaon photoproduction on the nucleon 
$\gamma p\to K^+\Lambda$ have been analyzed by means 
of a multipoles model. Different from the previous models, in 
this analysis the resonance decay widths are constrained
to the values given by the Particle Data Group (PDG). 
The result indicates that constraining these parameters 
to the PDG values could dramatically change the conclusion
of the important resonances in this reaction found in the
previous studies.
\keywords{Kaon; photoproduction; coupling constants.}
\end{abstract}

\ccode{PACS Nos.: 13.60.Le, 25.20.Lj, 14.20.Gk}

\vspace{2mm}


One of the most intensively studied topics in the realm 
of hadronic physics is kaon photoproduction. In the last 
decades a large number of attempts have been devoted to
model this reaction. Since this process is not dominated 
by any single resonant state, the main difference 
among these models is chiefly in the use of nucleon,
hyperon, and kaon resonances. Recently, a large number
of experimental data with good quality have been provided
by the SAPHIR\,\cite{SP03},  CLAS\,\cite{CL05},
LEPS\,\cite{Sumihama:2005er}, and GRAAL\,\cite{graal07}
collaborations. However, the lack of mutual consistency 
between the {\small SAPHIR} and other data found
by the recent phenomenological studies has increased
the difficulties of 
the extraction of the ``missing resonance'' properties.
In our previous work we investigated the physics 
consequence of using each data set\,\cite{Mart:2006dk}. It was found that
the use of SAPHIR and CLAS data, individually or simultaneously,
leads to quite different resonance parameters which, therefore, could
lead to different conclusions on ``missing resonances''.
In this paper we extend this investigation by constraining
the resonance decay widths to the values given by the
Particle Data Group\,\cite{pdg}. This is intended 
to approximately account for unitarity corrections
at tree-level, i.e., constraining the model by including some 
information from the leading $\pi$ and $\eta$ channels.

The background amplitudes of the model are constructed from a series 
of tree-level Feynman diagrams, consisting of the standard 
$s$-, $u$-, and $t$-channel Born terms along with the 
$K^*(892)$ and $K_1(1270)$ $t$-channel vector mesons. 
The resonant electric ($E_{\ell\pm}$), magnetic ($M_{\ell\pm}$), 
and scalar ($S_{\ell\pm}$) multipoles for a state with the mass $M_R$, width 
$\Gamma_R$ are assumed to have the Breit-Wigner form
\cite{hanstein99,Tiator:2003uu}
\begin{eqnarray}
  \label{eq:em_multipole}
  A_{\ell\pm}^R(W,Q^2) &=& {\bar A}_{\ell\pm}^R(Q^2) \, c_{KY}\, 
  \frac{f_{\gamma R}(W)\, \Gamma_{\rm tot}(W) M_R\, 
    f_{K R}(W)}{M_R^2-W^2-iM_R\Gamma_{\rm tot}(W)} ~,
  \label{eq:m_multipole}
\end{eqnarray}
where $W$ represents the total c.m. energy, $\ell$ indicates the kaon angular 
momentum, and $\ell\pm\equiv \ell\pm 1/2=j$ shows the total angular momentum. 
The isospin factor 
$c_{KY}$ is $\sqrt{3/2}$ and $-1/\sqrt{3}$ for the isospin
3/2 and isospin 1/2 \cite{hanstein99,Tiator:2003uu}, respectively, 
The factor $f_{KR}$ is the usual Breit-Wigner factor describing the 
decay of a resonance $R$ with a total width $\Gamma_{\rm tot}(W)$ 
and physical mass $M_R$, whereas $f_{\gamma R}$ indicates the 
$\gamma NR$ vertex.


\begin{figure}[!t]
\centerline{\psfig{file=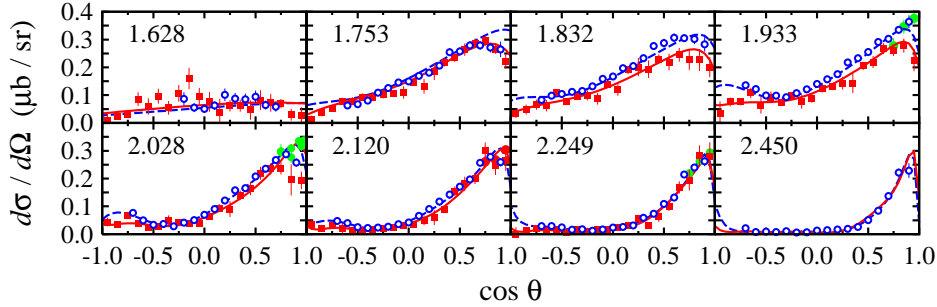,width=5in}}
\vspace*{8pt}
\caption{Comparison between angular distribution of differential 
  cross sections obtained from the two fits with CLAS (open circles) 
  SAPHIR (solid squares) and LEPS 
  (solid circles) data.
  \protect\label{fig:dcs}}
\end{figure}

\begin{figure}[!t]
\centerline{\psfig{file=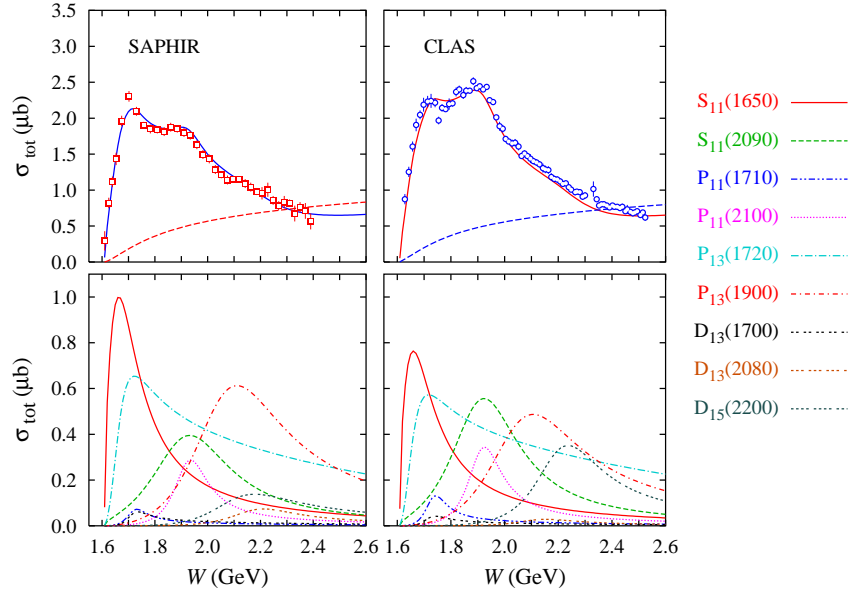,width=4.5in}}
\vspace*{8pt}
\caption{(Upper panels) Comparison between the calculated total cross sections
  (solid curves), contributions from background terms (dashed curves), and 
  experimental data. (Lower panels)
  Contributions from important resonances to the total cross sections.
  \protect\label{fig:total}}
\end{figure}

The results of fitting to the SAPHIR or CLAS data, 
compared with these data, are displayed
in Fig. \ref{fig:dcs}. Obviously, the 
model that fits to the SAPHIR data cannot 
perfectly explain the CLAS data, and vice versa. 
It also appears from this figure that the two data sets show the
largest discrepancy at $W\approx 1.9$ GeV in the forward and backward
directions. Consequently, at this energy the total cross section 
data show a bump (see the upper panels of Fig.~\ref{fig:total}), 
which corresponds to the ``missing resonance'' $D_{13}(2080)$
found in Ref.\,\cite{Mart:1999ed}. As shown in the upper panels 
of Fig.~\ref{fig:total} the CLAS data show a relatively larger
bump compared to the SAPHIR data. Clearly, this implies that
the extracted information on the responsible resonances for 
this peak could be different if we used different data sets.

Contributions of the background terms and the most important 
resonances are also shown in Fig.~\ref{fig:total}. Near the
production threshold the two well established resonances
$S_{11}(1650)$ and $P_{13}(1720)$ show important roles 
in both fits. This finding emphasizes the results of our 
previous studies\,\cite{Mart:2006dk,Mart:2008ik}. 
Reference\,\cite{Julia-Diaz:2006is} has also arrived at 
the same result, except for the $P_{13}(1720)$. Although
not too significant, Ref.\,\cite{Julia-Diaz:2006is} found 
that this state is still required in the case of SAPHIR
data. Comparison between the two dash-dotted curves in
the lower panels of Fig.\,\ref{fig:total} also confirms 
that the role of this state is more substantial in the
case of SAPHIR data. 

Another important state is the $P_{13}(1900)$, a two-star
resonance with the total width $\Gamma\approx 500$ MeV. Although
we found that the extracted mass is much larger than 1900 MeV,
its significant role found in all recent 
studies\,\cite{Mart:2006dk,Mart:2008ik,Julia-Diaz:2006is}
is also observed in the present analysis. On the other hand, the
resonance with the same quantum numbers but lower mass
$P_{11}(1710)$ plays insignificant role in this process,
which is consistent with the finding of our previous 
work\,\cite{Mart:2006dk}.

Compared to the previous studies, the only different result 
exhibited in Fig.\,\ref{fig:total} 
is the origin of the second peak in the total cross section.
As clearly shown, this peak originates from the contributions
of the $S_{11}(2090)$ and $P_{11}(2100)$ states. Most of the
recent investigations found that this peak indicates a 
``missing'' $D_{13}$
resonance with a total width $\Gamma$ varies from 165 to 570 MeV
\cite{Mart:2006dk}. Clearly the effect of constraining the fitted
parameters is quite significant in this case. Therefore, 
it is quite important to address this issue in the 
future single channel analyses of kaon photoproduction.

\begin{figure}[!t]
\centerline{\psfig{file=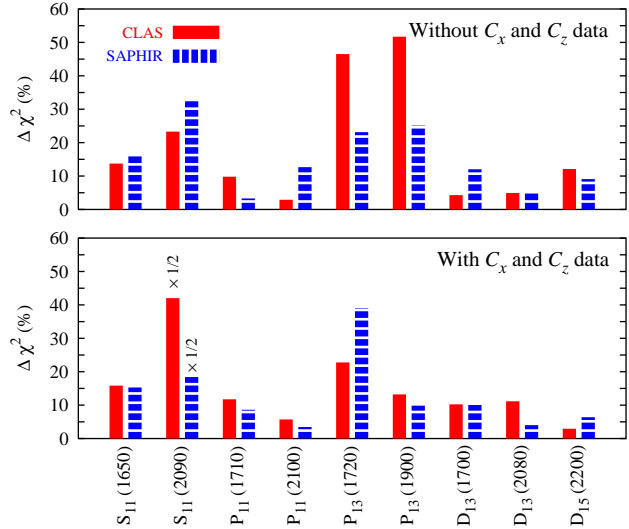,width=3.5in}}
\vspace*{8pt}
\caption{(Upper panel) The significance of individual resonances for
  fitting to the CLAS and SAPHIR data. (Lower panel) As in the upper
  panel, but including the CLAS $C_x$ and $C_z$ data.
\label{fig:strength}}
\end{figure}

It has been found that the inclusion of the new CLAS $C_x$ 
and $C_z$ data reveals the role of the $S_{11}(1650)$, 
$P_{11}(1710)$, $P_{13}(1720)$, and $P_{13}(1900)$ resonances 
for the description of these data\,\cite{Mart:2008ik}.
In this study we also investigate the effects of these data
on our model. The importance of the individual resonance for the
fits with and without these data is represented by $\Delta\chi^2
=|\chi^2_{\rm All}-\chi^2_{{\rm All}-N^*}|/\chi^2_{\rm All}\times 100\%$ in 
Fig.\,\ref{fig:strength}, where $\chi^2_{\rm All}$ is the $\chi^2$ 
obtained by using all resonances and $\chi^2_{{\rm All}-N^*}$
is the $\chi^2$ obtained by using all but a specific resonance.
Obviously, constraining the free parameters in the fits changes
the conclusion of the previous analyses. Only the  $P_{13}(1720)$ seems to be 
still important in both cases, whereas the  $P_{13}(1900)$ is only
important in the fit without $C_x$ 
and $C_z$ data. The near-threshold resonance $S_{11}(1650)$
is found to be relatively important in both cases. We also note
that in this study the $S_{11}(2090)$ 
is found to be quite important in both cases. The importance of this 
state has been actually found by the former study\,\cite{Mart:2006dk},
although with a smaller $\Delta\chi^2$ ($\approx 6\%$).

In conclusion we would like to say that the use of different data sets
could lead to different conclusions on the important resonances required
in the kaon photoproduction. Furthermore, 
constraining the free 
parameters in the multipoles model for this process results
in a significant effect and could dramatically change the
conclusion found in the previous analyses.
A more detailed study is currently
underway and the result will be published elsewhere in the near future.

The authors acknowledge the support from the University of Indonesia.


\begin{thebibliography}{0}
\bibitem{SP03} K.-H. Glander {\it et al.}, Eur. Phys. J. A {\bf 19}, 251
         (2004).
\bibitem{CL05} R. Bradford {\it et al.}, Phys. Rev. C {\bf 73}, 035202
         (2006).
\bibitem{Sumihama:2005er}
  M.~Sumihama {\it et al.},
  Phys.\ Rev.\ C {\bf 73}, 035214 (2006).
\bibitem{graal07}  A.~Lleres {\it et al.},
  Eur.\ Phys.\ J.\ A {\bf 31}, 79 (2007);
\bibitem{Mart:2006dk}
  T.~Mart and A.~Sulaksono,
  Phys.\ Rev.\ C {\bf 74} 055203 (2006).
\bibitem{pdg} C. Amsler {\it et al.}, Phys. Lett. {\bf 667}, 1 (2008).
\bibitem{hanstein99} D.~Drechsel, O.~Hanstein, S.~S.~Kamalov 
  and L.~Tiator, Nucl.\ Phys.\ A {\bf 645}, 145 (1999).
\bibitem{Tiator:2003uu}
  L.~Tiator {\it et al.},
  Eur.\ Phys.\ J.\ A {\bf 19}, 55 (2004).
\bibitem{Mart:1999ed}
  T.~Mart and C.~Bennhold, Phys.\ Rev.\ C {\bf 61}, 012201 (2000).
\bibitem{Mart:2008ik}
  T.~Mart, Few-Body Syst. {\bf 42}, 125 (2008).
\bibitem{Julia-Diaz:2006is}
  B.~Julia-Diaz, B.~Saghai, T.~S.~Lee and F.~Tabakin, Phys.\ Rev.\ C 
  {\bf 73}, 055204 (2006).
\end{thebibliography}
\end{document}